\documentclass[aps,superscriptaddress,onecolumn,floatfix,nofootinbib,preprintnumbers,12pt]{revtex4}
\usepackage{amsmath,multirow,graphicx,tabularx,epsfig}
\usepackage{tikz}
\usepackage{hyperref}
\usetikzlibrary{matrix,shapes.misc}

\newcommand\one{\leavevmode\hbox{\small1\normalsize\kern-.33em1}}

\def\k{\kappa}
\def\l{\lambda}

\def\e{{\rm e}}

\def\slashchar#1{\setbox0=\hbox{$#1$}           
   \dimen0=\wd0                                 
   \setbox1=\hbox{/} \dimen1=\wd1               
   \ifdim\dimen0>\dimen1                        
      \rlap{\hbox to \dimen0{\hfil/\hfil}}      
      #1                                        
   \else                                        
      \rlap{\hbox to \dimen1{\hfil$#1$\hfil}}   
      /                                         
   \fi}

\def\eg{{\sl e.g.} \,}
\def\ie{{\sl i.e.} \,}

\newcommand{\be}{\begin{eqnarray*}}
\newcommand{\ee}{\end{eqnarray*}}

\newcommand{\bee}{\begin{eqnarray}}
\newcommand{\eee}{\end{eqnarray}}
\newcommand{\beeq}{\begin{equation}}
\newcommand{\eeeq}{\end{equation}}

\def\eq#1{(\ref{#1})}

\def\fig#1{{Fig. \ref{#1}}}
\def\sec#1{{Section \ref{#1}}}
\def\ope{{\mathcal{O}}}

\begin{document}


\date{\today}

\title{Recursive prescription for logarithmic jet rate coefficients}

\author{Erik Gerwick}
\email{Erik.Gerwick@phys.uni-goettingen.de}
\affiliation{II. Physikalisches Institut, Universit\"at G\"ottingen, Germany}

\begin{abstract}
We derive a recursion relation for the analytic leading logarithmic 
coefficients of a final state gluon cascade.  We demonstrate the 
potential of our method by analytically computing the rate coefficients 
for the emission of up to 80 gluons in both the exclusive-$k_t$ (Durham) 
and generalized 
inclusive-$k_t$ class of jet algorithms.  There is a particularly simple form 
for the ratios of resolved coefficients.   We suggest potential applications 
for our method including the efficient generation of shower histories.
\end{abstract}

\maketitle


\section{Introduction}

Recursive algorithms are often the most efficient technique 
for calculating gauge theory amplitudes, as ideally information 
is maximally recycled \cite{Britto:2004ap,Berends:1987me,Parke:1986gb}.  
In recent years recursive techniques have become a major component for 
event simulation at the LHC, for tree-level generation 
of multi-jet events and part of the vast improvement in 
NLO calculations at higher multiplicity 
\cite{Gleisberg:2008fv,Badger:2010nx,
Bern:1994cg,Berger:2008sj}.  The 
irreducible complexity of full-matrix 
elements limit computations of final-state partons to a fairly modest number 
(typically  $n \le 10$ at LO, $n \le 5$ at NLO), which 
in the hard and widely separated regime 
meets essential experimental demand \cite{Berger:2010zx,Ita:2011wn,Aad:2013ysa}.  
However, for the logarithmically enhanced sector of soft and collinear radiation, 
generating high multiplicity is crucial and in practice proceeds through parton shower 
MonteCarlo~\cite{pythia,herwig,sherpa}. 

This paper introduces a simple technique for recursively 
extracting logarithmic coefficients of $n$-jet 
rates.  We emphasize that these coefficients are a mere 
skeleton of the complete (even tree-level matrix element) calculation, but 
our goal here is to explore the high multiplicity regime.  
We find simple implementations in both the 
exclusive-$k_t$ (here on, \emph{Durham}) 
and generalized inclusive-$k_t$ (here on, \emph{Generalized-$k_t$}) jet algorithms 
starting from their respective generating functionals  
\cite{fastjet,Catani:1991hj,Leder:1996py,Brown:1991hx,Webber:2010vz,Gerwick:2012fw}.  
The rates we calculate correspond to expanding in powers of $(\alpha_s /\pi) L^2$, where 
in the Durham algorithm $L$ is the logarithm of a dimensionless resolution scale 
$y_{cut}$, while in the Generalized-$k_t$ algorithm $L^2$ contains separate 
energy and angular logarithms which depends on a minumum energy scale 
$E_R$ and jet radius $R$.   It is known that the resolved coefficients obtained 
in this way are present in the LO matrix element calculation \cite{Brown:1991hx}, while the 
unresolved ones start at the NLO.  As we will see, since our formula allows the efficient 
computation of an exclusive $n$-gluon rate to arbitrarily 
high order in $(\alpha_s /\pi)L^2$ (\ie including additional 
unresolved gluons), for all practice purposes these rates can 
be thought of as resummed containing the same level of 
formal accuracy as a standard parton shower\footnote{
our implementation is for the double-leading-logarithms 
only, but the extension to the relevant next-to-double-logarithms 
also including the $g\to q\bar{q}$ splitting is discussed at a later stage.}.  It 
is important to bear in mind however that the rate coefficients do 
not \emph{a priori} contain 
any notion of kinematics or recoil as in the parton shower.

There are several potential applications for our work, all generally 
following from the ability to compute analytic expressions in a shower-like 
approximation.  To illustrate the improvement 
with an example, let us outline how the calculation proceeds directly 
from the generating functional for the exclusive rates in 
$e^+ e^- \to \bar{q}q + 20 \; \text{gluons}$ 
in the Durham algorithm.  Starting from the generating 
functional
\begin{alignat}{5}
\Phi_{g/q}(u,Q^2) &= u \; \exp 
\left[ \int_{Q_0^2}^{Q^2} dt \; \Gamma_{g/q}(Q^2,t) 
       \left( \Phi_g(u,t) -1 \right) \right] 
\label{eq:gf_evolution}
\end{alignat}
we obtain the resummed rate differentiating $(\Phi_q)^2$ 22 times 
with respect to the variable $u$ at the point $u=0$.  Thus we define 
the exclusive jet fractions 
 \begin{alignat}{5}
  f_{n} = \left. \frac{1}{n!} \frac{d^n}{du^n} \Phi_q^2 \right|_{u=0} .
\label{eq:def_gf}
\end{alignat}
The resulting resummed $20$-gluon expression we 
mercifully do not include, but note that it is a linear combination 
of 39,289,183 possible splitting histories\footnote{The number of splitting 
histories contributing to an $n$ gluon final state is the recursive number of 
integer sub-partitions of the integer partitions of $n$.}.  Obtaining a 
numerical answer requires either a numerical evaluation of each of the 
19-dimensional integrals (38 dimensional for the generalized 
class of $k_t$ algorithms) or for the fixed order coefficient, expanding 
the Sudakov form factors to the appropriate order and evaluating 
the still 19-dimensional integral analytically.  In practice this procedure could 
be optimized so that, for example, partial results for  
the multi-dimensional integrals are recycled, 
but it should be clear that the manipulations are extremely unwieldy.  

Expressing the expanded rates as the resolved and unresolved 
coefficients 
\begin{equation}
P_n = \text{Res}_n  \; + \; \text{URes}_n
\end{equation}
where $\text{Res}_n \sim \alpha_s^n $ and $\text{URes}_n$ starts 
at $\ope(\alpha_s^{n+1})$, our method allows the computation of 
$\text{Res}_{20}$ in a matter of seconds.  Once $\text{Res}_{20}$ is 
known, it is straight-forward to ``bootstrap" the unresolved 
components for the lower multiplcities using 
simple identical boson (Poisson) statistics.  Doing this to 
sufficiently high order, one recovers the resummed rates\footnote{
We note here very explicitly that the physics in our recursive 
prescription is identical to the coherent branching formalism.  In fact, we prove 
the consistency of our method directly from the generating functional.  What is 
special is that a simplified recursive formula allows us to study gluonic coefficients 
for arbitrary multiplicities, in practice an order of magnitude larger than using 
conventional techniques. 
}.  

The reason we are able to construct a simple recursive formula 
comes down to a well known fact about the exponentiation of 
leading singularities in gauge theory amplitudes, namely that it is 
determined by the maximally non-abelian contribution 
\cite{Frenkel:1984pz,Gatheral:1983cz} (for more recent results 
along these lines see \eg Refs.~\cite{DelDuca:2011ae,Gardi:2013ita}).  
For our prescription, which determines the coefficients of the 
leading soft-collinear singularities in the $L\to \infty$ sense, the only required 
physics input is the (coherent branching formalism analogous) maximally 
secondary coefficient, corresponding to a string of gluons each 
emitting exactly once.  This is diagrammatically encapsulated in the 
first moment of the generating functional \eq{eq:gf_evolution}, and these 
contributions are also order by order guaranteed to exponentiate.  Knowing 
only this contribution, the remainder 
of our recursive formula determines the entire leading coefficient 
using bosonic statistics.   We hope that our proof of the 
recursive algorithm makes this point clear.

%
%

This paper is arranged as follows.  In \sec{sec:Rec} we introduce 
the details of our recursive prescription for the resolved component.  
For the sake of presentation we prove the individual steps 
only at the end of the section.  We outline the 
method for pure Yang-Mills in the Durham algorithm.  
At the stated level of accuracy it is simple to 
generalize to arbitrary numbers of initial quarks or gluons.  
We include the prescription for the unresolved component 
in \sec{sec:URes}.  In \sec{sub_ex} we provide an example 
step in the recursion for $4$-gluon emission from a $q\bar{q}$ dipole.
In \sec{sec:genkt} we summarize the small 
modifications necessary for the 
inclusive-$k_t$ algorithm.  In \sec{sec:proof} we provide proofs for the 
individual steps of the recursion directly from the generating 
functionals.  We study the 
gluonic coefficients at high multiplicity in \sec{sec:app} and discuss some 
possible applications for our computational tool.  In the appendix 
we provide the resummed 6-gluon $f_6$ contribution used to validate our algorithm.

\medskip

\section{Recursive prescription}
\label{Recsas}
\subsection{Resolved Component}
\label{sec:Rec}

We consider here pure Yang-Mills (YM) in the Durham 
algorithm and start by decomposing the $n$-gluon final 
state in terms of its \emph{splitting history}.  We differentiate these from 
Feynman diagrams by distinguishing between the emitter and 
emitted parton at each $1\to2$ splitting.  We call each splitting 
involving an initial parton \emph{primary}, and any non-primary 
splitting is termed \emph{secondary}.
For fixed $n$ we write the resolved component of the 
corresponding $n$-gluon rate from a single initiator as   
\begin{equation}
\text{Res}_n \;\; = \;\; \sum_k \sum_{i=0}^{n-1} c^{(n)}_{ik} 
\left(a_s C_A L^{2} \right)^n 
\label{defers}
\end{equation}
where $a_s = \alpha_S/\pi$, $L = \log(1/y_{\text{cut}})$ and  
$c^{(n)}_{ik} > 0$.  The index $i$ counts the number of secondary 
emission in a particular splitting history.  
The sum on $k$ is over all diagrams of the same order in $i$, which 
is left implicit for the moment.  Our definition ensures that every term in 
\eq{defers} is in one-to-one correspondence with a specific 
splitting history.  However, the recursive formula for the 
resolved coefficients does not depend on the index $k$, so 
we drop it for the time being.  

We claim that given a specific subset of coefficients from multiplicities $n$ and smaller, we 
can write a general expression for $c^{(n+1)}_{i}$, and using \eq{defers}, 
compute $\text{Res}_{n+1}$.   The necessary ingredients 
for $c^{(n+1)}_{i}$ are:
\begin{itemize}
\item All of the $c^{(l)}_{l-1}$ with $l < n+1$.  These 
are all of the previous coefficients highest order in secondary 
emissions, or in other 
words containing precisely one primary splitting from the initial 
hard line.  
\item All of the $c^{(n)}_{l-1}$ with $l -1< n$.  These 
are the coefficients with at least two primary splittings only 
from the $n$-th coefficient.  
\item Integer partitions of $n+1$.\end{itemize}
Using these ingredients we find a simple formula for the rate coefficients.  
To illustrate this procedure we first go through the steps in the recursive 
prescription.  We provide a detailed example of one step in the recursion 
for 4-gluon emission in Section. \ref{sub_ex}.

\begin{figure}[t]
\includegraphics[width=1.0\textwidth]{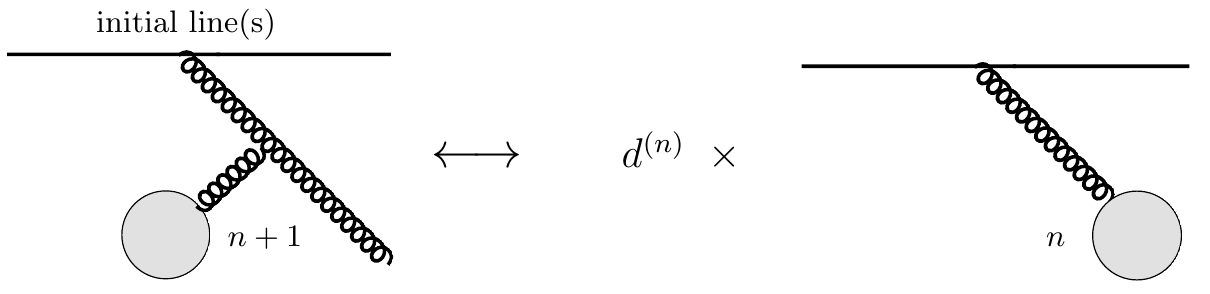}
\caption{Illustration of the first term in \eq{easlif}.  The solid line always represents 
the initial partons in the process, which for our current example is a single gluon.}
\label{ill3}
\end{figure}

The first step is to divide the coefficients into two categories
\begin{equation}
c^{(n)}_{i} = c^{(n)}_{k} + c^{(n)}_{n-1}
\end{equation}
The index $i \in (0,n-1)$, $k \in (0,n-2)$.
The $c^{(n)}_{k}$ coefficients are the contributions with at least 2 primary splittings.  
Each gluonic structure is already present in the lower multiplcity coefficients, 
and can therefore be constructed by multiplying such coefficients and taking into 
account symmetry factors (this is intuitive, although it is proven in 
Sec. \ref{sec:proof} more explicitly).
In contrast, the $c^{(n)}_{n-1}$ coefficients are maximally secondary with respect 
to the hard initial line.  These satisfy a relatively simple recursion relation for 
promoting coefficients higher up on the emission tree
\begin{equation}
c^{(n+1)}_{n} \; =\;   \; \sum_{j=0}^{n-1} c^{(n)}_{j}  d^{(n)} 
\qquad \qquad d^{(n)} = \dfrac{(2n)!}{(2n+2)!} \; .
\label{easlif}
\end{equation}
Diagrammatically the $n-1$ term in \eq{easlif} corresponds to the relation in 
Fig. \ref{ill3}.
The grey blob indicates that this gluon is allowed to emit an arbitrary 
number of times, and each emission itself may split \emph{et cetera}.  
The solid line will always indicate an arbitrary number and type of initial 
partons, which for this specific example we take as a single gluon.
The other terms in \eq{easlif} sums over the $c^{(n)}_j$ terms not maximally 
secondary and not representable in the relation above.
We see that the two step process promotes diagrams with at least two  
primary emissions to ones on the RHS and finally to the LHS of Fig. \ref{ill3}.  
The origin of the specific form of \eq{easlif} is that the prescription for 
promoting primary to secondary emission essentially involves reweighting
by the first moment of the generating functional, which for the Durham 
algorithm is $\Phi'_{u=1} \sim \sum_{n=0}^{\infty} (aC_A L^2)^{2n}/(2n)!$ \cite{Ellis:1991qj}.  
Diagrammatically, this is identical to the sum  of maximally secondary 
splitting histories.

The final step in our recursion is to generate the $c^{(n+1)}_k$ with $k < n$ 
coefficients.  It is easy to see that a recursion based solely on $c^{(n)}$ 
coefficients is bound to fail, as the integer partition of $n$ arising at each multiplicity 
is not easily defined recursively.  Instead, we compute $c^{(n+1)}_k$ by enumerating 
the various partitions of gluons and weighting by the appropriate irreducible 
structures $c^{(k)}_{k-1}$.  Note that only the values of $c^{(k)}_{k-1}$ need 
to be stored from previous multiplicities.  
Computing $n+1$ coefficients we only require $n$ of such numbers 
making this step computationally manageable\footnote{Looping over the 
various partitions still constitutes the most computationally intensive part 
of our algorithm.}.  An additional ingredient is that 
$m$ identical structures carries a phase space factor $1/m!$.  

\begin{figure}[t]
\includegraphics[width=1.0\textwidth]{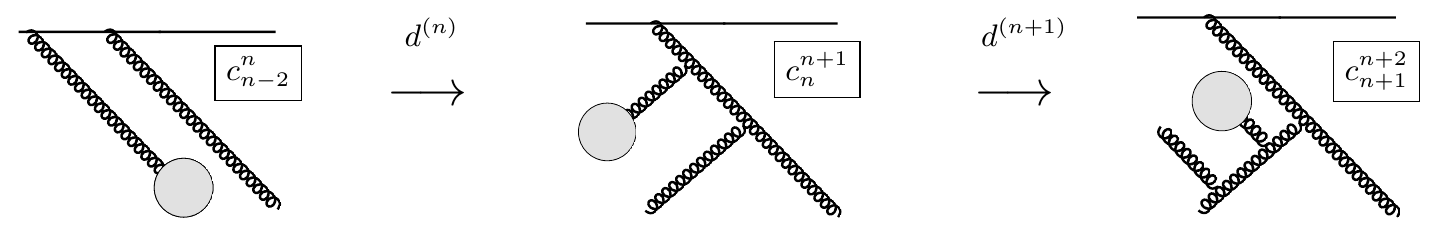}
\caption{Example recursion for an individual diagram.  As soon as 
the diagram ends up on the right-hand side it is repeated in the 
recursion according to the $d$ term in \eq{easlif}.}
\label{exred}
\end{figure}

A complete representation for this contribution is
\begin{equation}
c^{(n+1)}_{k} =  \sum_{p(n)} \frac{1}{S} \left[ \prod_{\sigma_i=\{ \sigma_1,\cdots\, \sigma_{r}\} } \;
c^{(\sigma_i)}_{\sigma_i-1}\right].
\label{c1rec}
\end{equation}
where the sum is over integer partitions $p(n)$ of $n$ of length 
$r \ge 2$.  The product is over the individual elements of each 
partition.  For example, for $n = 4$ there are 4 partitions in the 
sum $\left\{\sigma_1,\sigma_2,\sigma_3,\sigma_4\right\} = 
\{(3,1),(2,2), (2,1,1),(1,1,1,1)\}$. 
Here $S$ is the overall symmetry number taken as the 
product of identical structure phase space factors, \eg the 
contribution from the (2,2) term is 
$(1/2!)(c^{(2)}_{1})^2$.
We can summarize the entire recursive algorithm for the resolved 
coefficients and the main result of this paper 
\begin{equation}
\sum_{i=0}^{n} c^{(n+1)}_{i}  = \sum_{p(n)} \frac{1}{S} \left[ \prod_{\sigma_i=\{ \sigma_1,\cdots\, \sigma_{r}\} } \; 
c^{(\sigma_i)}_{\sigma_i-1}\right]
\;\; + \;\;\sum_{j=0}^{n-1} c^{(n)}_{j}  d^{(n)} 
\label{final}
\end{equation}
An example recursion for an individual diagram is given in Fig. \ref{exred}.  Note 
that as soon as a diagram ends up in the furthest right $c^{(n+2)}_{n+1}$ class it 
remains there indefinitely.  It is simple to check that our formula exhausts all 
possible splitting histories to a given multiplicity.  We confirm the validity 
of \eq{final} by comparing with a direct computation from the generating 
functional with up to $5$ final state gluons \cite{Gerwick:2012hq}.

\bigskip

\subsection{Unresolved Component}
\label{sec:URes}

Given the set of resolved coefficients up to multiplicity $n$, it is 
relatively straight-forward to determine the unresolved coefficients 
for lower multiplicitities also up to order $(a_s L^2)^n$.  To describe these 
coefficients we extend our notation slightly so that $c^{(l)}_{i} \to c^{(l,n)}_{i}$ 
where $l$ ranges between $0, 1,\cdots n$ and indicates the 
multiplicity.  The resolved coefficients are then $c^{(n,n)}_{i}$ 
and the unresolved are the rest.  Now it should be clear that 
the unresolved coefficients 
come from expanding the Sudakovs beyond leading order.  
Therefore, we expect the unresolved coefficients to be related to an expanded 
exponential and most importantly, to be determined from the resolved 
components at the same order.

For the simplest case of the all primary contributions we find
\begin{equation}
c^{(l,n)}_{0} = (-1)^{n-l} \frac{1}{(n-l)!} \frac{1}{l!} \;.
\label{simp_res}
\end{equation}
Note that at every order the individual coefficients correctly 
satisfy $\sum_{l=0}^{n} c^{(l,n)}_{0}  = 0$.  This fact holds on 
a diagram by diagram basis to all multiplicities for the exclusive rates.  
In order to extend also to the secondary terms, the complication is that we 
need to distinguish diagrams beyond what we have so far for the resolved 
component.  The additional necessary ingredient is the number of 
repeated identical emissions in a given splitting history. 

In order to proceed, let us note that due to our recursion relation 
the resolved component $c^{(n,n)}_{j}$ of each splitting history is known, 
and can be decomposed in terms of numerical coefficients times 
powers of $c^{(1,1)}_{0}$, $c^{(2,2)}_{0}$ and $c^{(2,2)}_{1}$.  
These are our starting conditions, which we will refer to as the \emph{primordial 
coefficients}.  Let us denote the powers of each as $a$, $b$ and $c$ 
respectively.  Now we define 
\begin{equation}
p = a + 2b + c
\label{nuaasp}
\end{equation}
and claim that for the unresolved components $c^{(l,n)}_{j}$ of this 
particular diagram, that $l \in (n, n-1, \cdots , n-p)$ with coefficients 
given by 
\begin{equation}
c^{(l,n)}_{j} = (-1)^{n-l}  \frac{1}{(n-l)!}\frac{ p! }  {(p-(n-l))!} c^{(n,n)}_{j} \, .
\label{fin_un_r}
\end{equation}
For $l<n-p$ we set $c^{(l,n)}_{j}=0$.
The resolved coefficients along with $p$ determine entirely the 
right-hand side of \eq{fin_un_r}.  Again, the physical interpretation 
of this formula suggests Poisson statistics.  Note that the maximally 
primary emission 
formula \eq{simp_res} is a special case of \eq{fin_un_r} with 
$c^{(n,n)}_j=c^{(n,n)}_0 = (1/n!)$ and $p = n$.  Also we 
recover the resolved coefficient when $n=l$.
In analogy to \eq{simp_res} each emission history separately 
obeys the unitarity condition $\sum_{l=0}^{n} c^{(l,n)}_{j}  = 0$.  

We say more about the specific terms in \eq{fin_un_r} in \sec{sec:proof}.  
For now we proceed through an explicit step in the recursion for $4$-final state 
gluons.

\subsection{Example calculation for 4-gluons in $e^+ e^-$}
\label{sub_ex}
We demonstrate our recursive prescription by generating 
the $4$-gluon final state in $e^+ e^- \to q\bar{q} + n$ gluons  
from the lower multiplicity coefficients.  This is defined as 
$f_6$ in the literature.  We choose this multiplicity because 
it is simple to check but includes the non-trivial features of our 
algorithm. 
First we note from Ref. \cite{Gerwick:2012fw} that at the stated 
level of accuracy our coefficients relate to those in $e^{+} e^{-} $ through
\begin{equation}
\Phi_{ee} \; = \Phi_q^2\;=\; u^2\,(\Phi_g/u)^{2\,C_F/C_A} \,.
\end{equation}
which suggests that we define the resolved coefficients in $e^+ e^-$ as 
\begin{equation}
\text{Res}_n^{(e^+ e^-)} \;\; = \;\; \sum_k \sum_{i=0}^{n-1} c^{(n)}_{ik}  
C_A^i C_F^{n-i} \left(a_s L^{2} \right)^n .
\label{eedefers}
\end{equation}
Therefore, in our recursion relation we have 
\begin{equation}
d^{(n)}_j = \dfrac{2^{j-n+1}}{(2n+1)(n+1)}\, ,
\label{new_j}
\end{equation}
which together with \eq{eedefers} provide the correct description
for $e^+ e^- \to q\bar{q} + n$ gluons.  The $j$ dependence 
in \eq{new_j} reflects the fact that a secondary emission necessarily 
comes off a single gluon, while there are two possible quark 
lines for a primary emission.
The initial values for the lower coefficients are
\begin{alignat}{5}
(c_{0}^{(1)}, c_{0}^{(2)} , c_{1}^{(2)},c_{0}^{(3)},c_{1}^{(3)},c_{2}^{(3)} ) = 
\left( 1 , \frac{1}{2} ,\frac{1}{12}, \frac{1}{6},\frac{1}{12},\frac{1}{90} \right).
\end{alignat}
The partitions of $n=4$ from \eq{c1rec} gives for the $c^{(4)}_i$ with $i\le2$
\begin{alignat}{5}
c^{(4)}_0 = \frac{1}{4!}\left(c_{0}^{(1)}\right)^4 \qquad
c^{(4)}_1 = \frac{1}{2!}\left(c_{0}^{(1)}\right)^2 c_{1}^{(2)}  \qquad
c^{(4)}_2 = \frac{1}{2!} \left( c_{1}^{(2)} \right)^2 \, + \,c_{0}^{(1)} c_{2}^{(3)}.
\label{easy1}
\end{alignat}
For the highest order terms in secondary splittings (in $C_A$ in this case) 
we have from~\eq{easlif}
\begin{alignat}{5}
c_{3}^{(4)} = \frac{1}{28}\left[ 2^3 c_{0}^{(3)}   
+ 2^2 c_{3}^{(1)} \right] + \frac{1}{56} c_2^{(3)},
\label{easy2}
\end{alignat}
so that we find
\begin{alignat}{5}
(c_{0}^{(4,4)} , c_{1}^{(4,4)}, c_{2}^{(4,4)},c_{3}^{(4,4)} ) = 
\left( \frac{1}{24} , \frac{1}{24} ,\frac{7}{480}, \frac{17}{10080} \right)\, ,
\label{res_6}
\end{alignat}
where we have emphasized that these are the resolved coefficients by 
adding the label indicating the power of $a_s L^2$.
Possessing the resolved coefficients we now compute the unresolved 
ones.  Expressing \eq{easy1} and \eq{easy2} above in terms of their primordial  
coefficients we find
\begin{alignat}{5}
c^{(4,4)}_0 &= \frac{1}{4!}\left(c_{0}^{(1)}\right)^4 \\
c^{(4,4)}_1 &= \frac{1}{2!}\left(c_{0}^{(1)}\right)^2 c_{1}^{(2)}  \\
c^{(4,4)}_2 &= \frac{1}{2!} \left( c_{1}^{(2)} \right)^2 \, + \,c_{0}^{(1)} \left(
\frac{1}{2^2\cdot15} c_{0}^{(2)} + \frac{1}{30} c_{1}^{(2)} \right)\, \\
c_{3}^{(4.4)} &= \frac{1}{28}\left[ \frac{1}{2^3\cdot 3!} (c_{0}^{(1)})^3   
+ \frac{1}{2^2} c_{0}^{(1)} c_{1}^{(2)} \right] + \frac{1}{56} \left(
\frac{1}{2^2\cdot15} c_{0}^{(2)} + \frac{1}{30} c_{1}^{(2)} \right).
\label{easym2}
\end{alignat}
Each term in the above expressions is in one-to-one correspondence with 
a specific emission history (there are 9 at this multiplicity).  Applying 
\eq{fin_un_r} we find the unresolved set of coefficients  
\allowdisplaybreaks
\begin{alignat}{5}
& \left( c^{0,4}_0 \right) \; =
\left( \frac{1}{24}\right) 
\notag \\
& \left( c^{1,4}_0,c^{1,4}_1,c^{1,4}_2,c^{1,4}_3 \right) \; = 
\left(-\frac{1}{6}, - \frac{1}{24},-\frac{1}{120},-\frac{1}{1344} \right)
\notag \\
& \left( c^{2,4}_0,c^{2,4}_1,c^{2,4}_2,c^{2,4}_3 \right) \; =
\left(\frac{1}{4},\frac{1}{8},\frac{1}{32},\frac{1}{320} \right)
\notag \\
& \left(c^{3,4}_0,c^{3,4}_1,c^{3,4}_2, c^{3,4}_3 \right) 
 \; = 
\left(-\frac{1}{6},-\frac{1}{8},-\frac{3}{80}, -\frac{41}{10080} \right)\, .
\label{unres_6}
\end{alignat}
We can easily check \eq{res_6} and \eq{unres_6} from the resummed 
expression for $4$ gluons emission (in the Appendix Eq.~\eq{f6}) by 
expanding the Sudakov form factors beyond leading order.

\subsection{Prescription for Generalized-$k_t$ algorithm}
\label{sec:genkt}
Our recursive prescription also provides 
a formula for the Generalized-$k_t$ class of jet algorithms.
As a starting point, we list the generating functional 
for this algorithm \cite{Gerwick:2012fw}
\begin{equation}\label{eq:PhigDLA}
\Phi_{g/q}(u,\k,\l) =  u\,\e^{-a_{g/q}\k\l}\exp\left\{a_{g}\int_0^\k d\k' \int_0^\l d\l'
\,\Phi_g(u,\k',\l')\right\}\;.
\end{equation}
We define the logarithmic variables $\k=\log(E/E_R)$ and 
$\l = \log(\xi/\xi_R)$, with $\xi = 1 - \cos \theta_{ij}$  and 
$\xi_R = 1 - \cos R$.  The scale $E$ is identified with the initial 
hard scale of the process and $E_R$ the cut-off (in hadron 
colliders the transverse momentum).  The angle $\theta_{ij}$ is 
the opening angle between the emitting and emitted parton.

In just the same way as before we can divide the splitting 
histories into two sets of coefficients, each with 
its own distinct diagrammatic class.  The first moment of 
the generating functional in this case is
\begin{equation}
\Phi'_{g}|_{u=0} \, = \, I_0\left(2\sqrt{C_Aa_s \k\l}\right) \, ,
\end{equation}
%
%
%
where $I_0$ is the Bessel function $I_0(x) = \sum_{n=0}^{\infty} x^n/(n!)^2$.  
From this we immediately find the coefficient for promoting emissions higher 
on the tree.  In analogy to ${d}^{(n)}$ we have
\begin{alignat}{5}
\tilde{d}^{(n)} & = \frac{(n!)^2}{([n+1]!)^2} \, .
\end{alignat}
The step of the recursion involving the symmetry factors is identical. The recursive 
prescription for the generalized $k_t$ algorithm is simply 
given by \eq{final} with ${d}^{(n)}$ replaced 
with $\tilde{d}^{(n)}$.  The initial values of 
the primordial coefficients also differ from the Durham algorithm.  We 
have confirmed the recursive prescription for the Generalized-$k_t$ algorithm 
by comparing with the direct computation for up to $6$ jets in $e^{+} e^{-}$.

\subsection{Proof of Recursive formula}
\label{sec:proof}

In this section we show the respective parts of the 
recursive program.  As discussed in the introduction, 
computing numerical coefficients from the generating 
functional formalism is a two step process, the 
differentiation with respect to $u$ (computation of the 
resummed rates) and the evaluation 
of the $k_t$ integrals.  Therefore, we first establish the 
recursive behavior of the differentiation and then 
analyze the structure of the resulting chain of integrals.   
\bigskip

\underline{$c^{(n+1)}_{n}$ step:} \quad
Here we prove the part of the recursion in \eq{easlif} for the 
coefficients highest order in secondary emissions.  These come 
from differentiating the generating functional in all 
combinations where precisely $1$-derivative acts 
on the first $u$ in the exponent.  From the definition 
of the rates we find  
\begin{equation}
c^{(n)}_{n-1}  \; \; \sim \; \; \frac{1\;}{(n-1)!}
 \int_{Q_0^2}^{Q^2} dt \; \Gamma_{g}(Q^2,t)
\; \left( \left[ \dfrac{d}{du} \right]^{n-1} \Phi_g(u,t) \right).
\label{cn23}
\end{equation}
On the other hand we find for the $c^{(n+1)}_n$ the 
following relations
\begin{alignat}{5}
c^{(n+1)}_{n} \; \;  \sim
\,\frac{n(n+1)}{(n+1)!}
  \int_{Q_0^2}^{Q^2} dt \; \Gamma_{g}(Q^2,t)
\; \int_{Q_0^2}^{t'} dt' \; \Gamma_{g}(t,t')
\; \left( \left[ \dfrac{d}{du} \right]^{n-1} \Phi_g(u,t)  \right).
\label{cn133}
\end{alignat}
The pre-factor in the numerator of \eq{cn133} comes from the fact
that $n+1$ and $n$ derivatives can bring down the 
first and second convoluted gluon respectively.  
We recognize that promoting the coefficients gives the same 
statistical pre-factors from the $u$ differentiation, 
although each integral now contains an additional integral at 
the top end (nearest the hard process).  This is 
easily found by substitution, so that \eq{cn23} and \eq{cn133} are 
related through
\begin{alignat}{5}
c^{(n+1)}_{n} \; & \sim \; c^{(n)}_{n-1} 
  \int_{Q_0^2}^{Q^2} dt \; \Gamma_{g}(Q^2,t)
  \log^{2n}\left( \frac{t}{Q_0^2} \right) \notag \\
	\; &= \; \frac{1}{(2n+2)(2n+1)} c^{(n)}_{n-1}
\label{prosse}
\end{alignat}
providing the result in \eq{easlif}.  We note that the pre-factors 
originate from the growth of the maximally secondary (fully convoluted) 
integral, and are precisely the coefficients from the 1st-moment of the 
generating functional.  For $k$ initial lines there is the additional 
factor counting the $j$ dependence $k^{j-n+1}$, for example reflected 
in \eq{new_j}.
\medskip

\underline{$c^{(n+1)}_j$ step:}  
We prove the second step \eq{c1rec} in the recursion. To see 
how this comes about note that the differentiation with respect to 
$u$ produces all integer partitions of $n+1$, so it is only a matter 
of understanding the pre-factors.  Let us start first with the case of 
$n-2$ primary emission and a single secondary splitting.
\begin{alignat}{5}
\vcenter{\hbox{\includegraphics[width=0.17\textwidth]{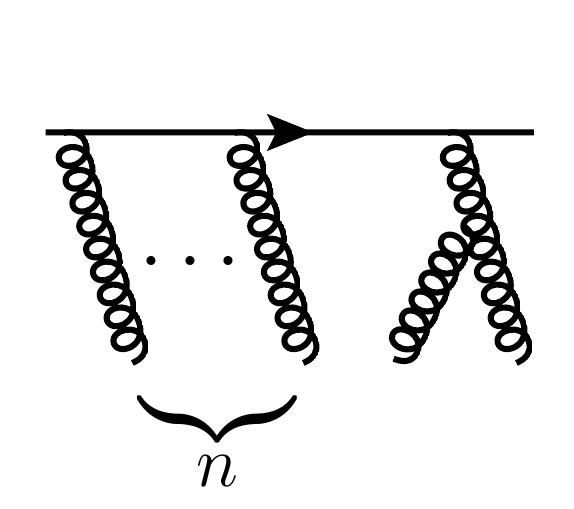}}}
\;\; = \;\; & \;\frac{1}{(n+1)!} \left[(n+1)\,\Gamma_q^{n-2}\right]\, \left[n(n-1) (\Gamma_q 
\otimes\Gamma_g)\right] \, \Phi_g|_{u=0}\notag \\
\;\; =  \;\;& \;\frac{1}{(n-2)!}\left( \Gamma_q\right)^{n-2} \, \left( 
\Gamma_q \otimes\Gamma_g \right) \,  
\end{alignat}
where the circle multiplication indicates that the upper integration limit on the gluon virtuality is 
set by the evolution of the quark.  From here we can read-off that 
$c^{n+1}_1 = (1/(n-2)!) \, c^{(2)}_{1}\, (c^{(1)}_{0})^{n-2}$.  In other words the 
composite coefficient is found by simply multiplying the exact terms from 
the previous irreducible structures. 

Now, let us consider a contribution corresponding to the generic partition of length 2, $\{ \sigma_1, 
\sigma_2\}$ of $n$.  Differentiating we find
\begin{alignat}{5}
\vcenter{\hbox{\includegraphics[width=0.17\textwidth]{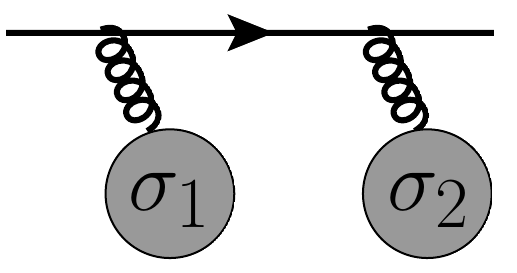}}}
&\;\;  = \;\; \frac{1}{(n+1)!}\left( \frac{d}{du} \right)^{\sigma_1+ \sigma_2 +1} 
\Phi(u) \notag \\
& \;\; =  \;\;\frac{1}{S} 
\frac{1}{\sigma_1! \, \sigma_2!}
\left( \frac{d}{du} \right)^{\sigma_1}
\left( \frac{d}{du} \right)^{\sigma_2} 
\Phi(u)  \left[ 1 + u \, f_1(u) +\cdots\right] |_{u=0}
\label{gen_proof}
\end{alignat}
where the notation indicates that we restrict ourselves in the 
second line to terms where the derivatives producing 
$\sigma_1$ and $\sigma_2$ act on separate ``$u$-trees".  We 
are permitted to consider only this contribution from 
\eq{gen_proof} since we include the compensating binomial 
coefficient 
\begin{equation}
\binom{n+1}{\sigma_1}  = \frac{(n+1)!}{\sigma_1! \,\sigma_2!} ,
\end{equation}
which give us precisely the denomenators needed to define the lower 
coefficients $c^{(\sigma_1)}_{\sigma_1-1}$ and $c^{(\sigma_2)}_{\sigma_1-2}$.
The symmetry factor is as previously defined so that for the partition under 
consideration, $S = 2!$ if $\sigma_1 = \sigma_2$ and $S=1$ 
otherwise.  This comes about because in that case, when the two 
$u$-trees are identical, we are over-counting the first 
differentiation in the above.  Thus we find for this particular example
\begin{equation}
c^{(n+1)}_{n-1} = \frac{1}{S} c^{(\sigma_1)}_{\sigma_1-1}
c^{(\sigma_2)}_{\sigma_2-1}.
\end{equation}
The generalization to arbitrarily complicated partitions 
of gluons follows directly through induction.

\underline{Unresolved components:}
Here we provide arguments for \eq{fin_un_r} for the unresolved 
components.  It is instructive to first point out the origin 
of the quantity $p$ defined in \eq{nuaasp}.  We consider stripping 
a splitting history of all emissions which are repeated off a single 
line, thus defining the \emph{stripped history}.  The stripped splitting 
history is always in the class of maximally secondary 
contributions (or products thereof), and has precisely $p = 1$.  
Now for a general history $p$ counts the number of repeating 
splittings.  The unresolved components of these splittings are 
found in the lower order multiplicity rates from the Sudakov expanded 
beyond leading order.  

In this language we can describe the various terms in \eq{fin_un_r}.  
The first reflects that these coefficients necessarily come from 
expanding the $\exp{(-\Gamma)}$ and the exponent $n-l$ counts 
the number of identical unresolved gluons to be divided 
over.  The next term contains the symmetry factor for the unresolved 
gluons, while the $(p-(n-l))!$ is for the remaining resolved gluons.  
The $p!$ in the numerator is the normalization required for unitarity.

\section{Applications and Discussion}
\label{sec:app}

\begin{figure}[t]
\centering
  \includegraphics[scale=0.87]{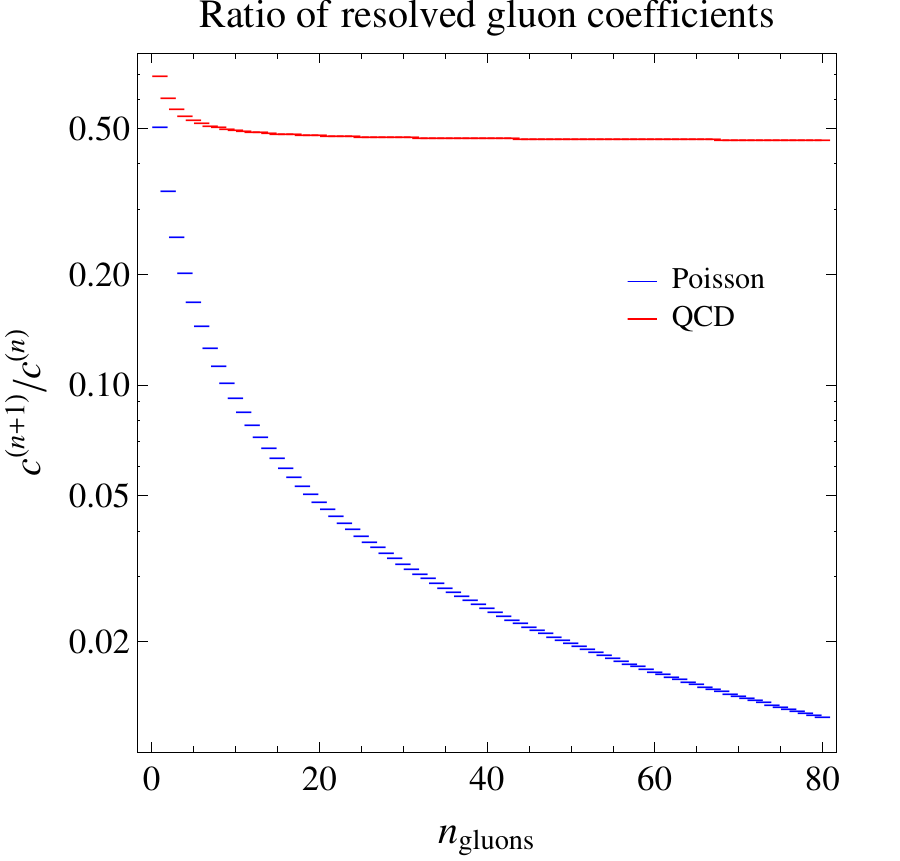}   
    \put(-170,40){\scriptsize Durham algorithm}
  \hfill
\includegraphics[scale=0.89]{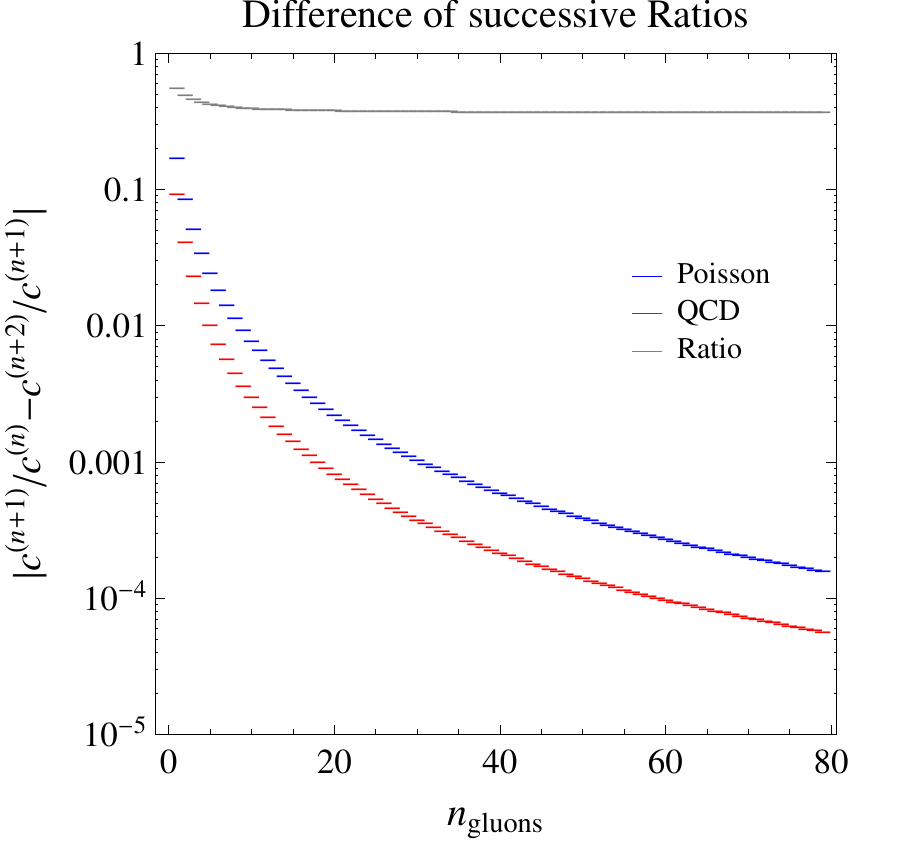}
\put(-170,40){\scriptsize Durham algorithm}
\caption{Left panel: Ratio of resolved (Durham algorithm) coefficients for up to $80$ gluons 
from a quark initiated gluon cascade compared with a Poisson radiation 
pattern.  Right panel: Difference between successive ratios from left panel 
showing the convergence to a constant like $1/n$.  In the case of the 
Poisson the ratios convergence to $0$ while for QCD they converge to 
the value $0.45622$. }
\label{fig:rat}
\end{figure}

\begin{figure}[t]
  \includegraphics[scale=1.0]{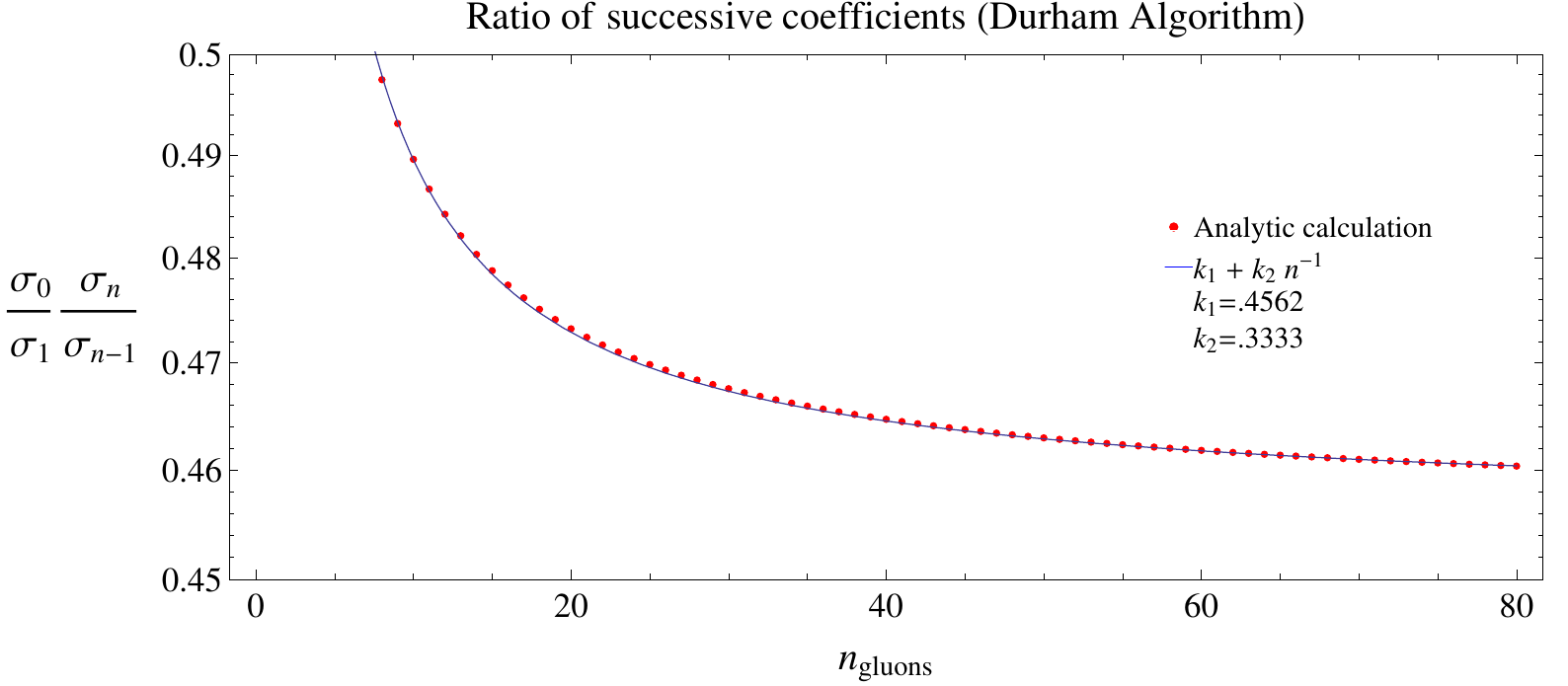}    
\caption{ Close-up of the QCD curve in the left panel of \fig{fig:rat} together with 
the fit function $k_1 + k_2/n$.  The red dots represent the (normalized) ratios 
of analytic leading logarithmic resolved coefficients for a pure gluon cascade. }
\label{fig:rat_fit}
\end{figure}

\underline{Ratios of resolved coefficients:} 
As a first application we study the ratio of 
resolved coefficients.  
In \fig{fig:rat} we show the ratios of successive 
exclusive coefficients.  We find an emerging geometric behavior 
in the ratio which dramatically contrasts a Poisson process 
(see left panel of \fig{fig:rat}).  However, we see from the 
right panel of \fig{fig:rat} that both approach 
their asymptotic limit with the functional dependence 
$(\text{constant})/n$.     The ratios are fit to remarkable precision with 
the function $c_{n+1}/c_n = k_1 + k_2/n$.  For Poisson 
we have the fairly 
uninteresting  $k_1=0$ and $k_2=1$ (we have normalized the first ratio to $1$ 
here).  For QCD in the Durham algorithm we 
find $k_1=0.4562$ and $k_2=0.3333$.  This fit is shown in \fig{fig:rat_fit}.  The 
remarkably simple form for the fit and the particular value 
of $k_2$ suggest an emergence from underlying 
dynamics.  
Calculating $k_1$ and $k_2$ from first principles 
starting either from the generating functional or some other 
form of QCD resummation would be interesting.

\begin{figure}[t]
\centering
  \includegraphics[scale=0.80]{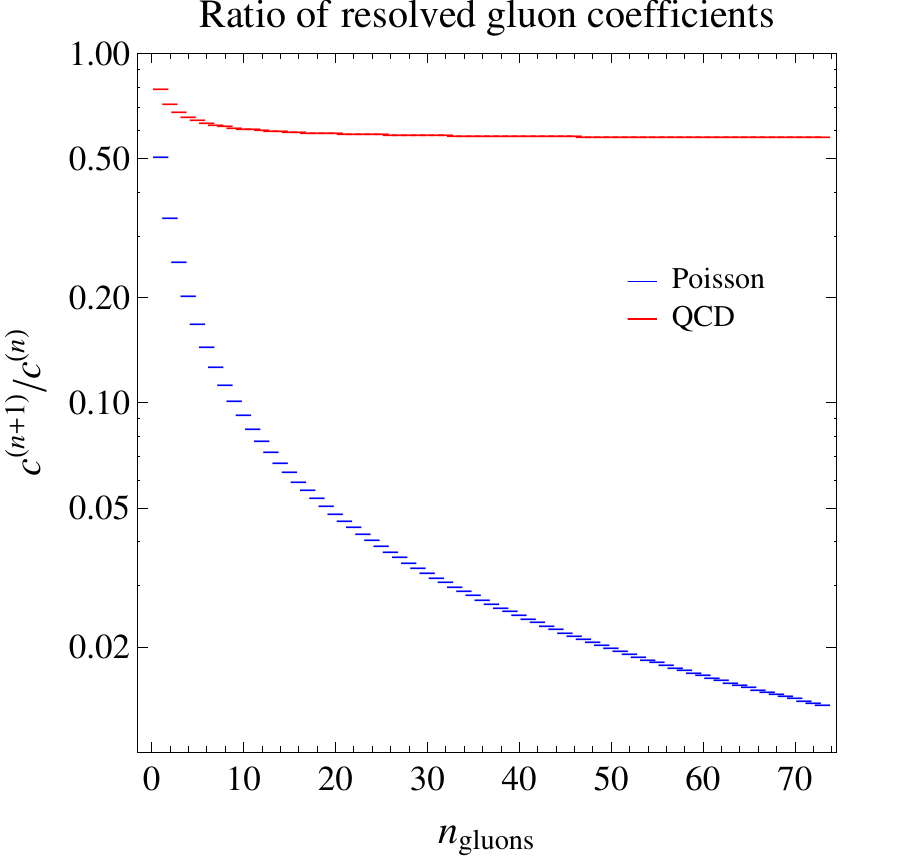}   
  \put(-160,40){\scriptsize Gen-$k_t$ algorithm}
  \hfill
\includegraphics[scale=0.82]{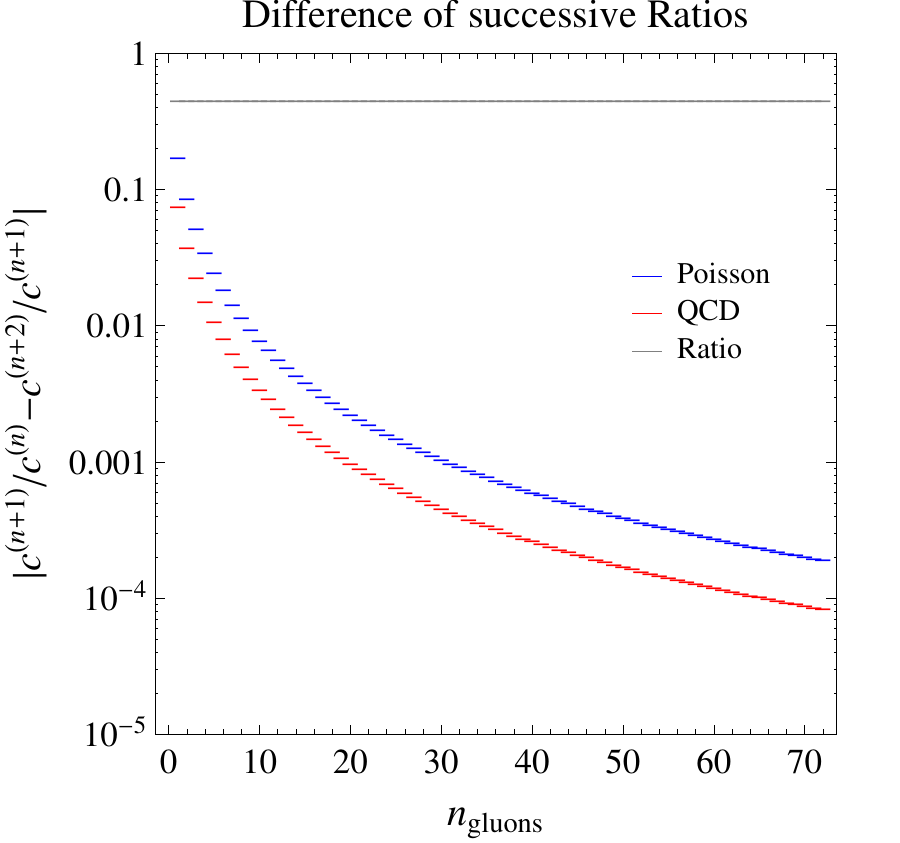}
  \put(-160,40){\scriptsize Gen-$k_t$ algorithm}
\caption{Same as Fig. \ref{fig:rat} but in the generalized 
$k_t$ algorithm.  Not that the right panel depicting 
the ratio of differences between successive ratios in the 
Poisson and generalized $k_t$ case is exactly constant.  }
\label{fig:rat_gen}
\end{figure}

In \fig{fig:rat_gen} we repeat the analysis of \fig{fig:rat} for 
the Generalized-$k_t$ recursive formula.  In this case 
we find an additional interesting feature, that the ratio 
between the difference of ratios for Poisson versus QCD 
is an exact constant $=\, 7/16$.  A full explanation for 
this behavior likely comes from number theory, since 
we see a relation between quantities built from integer 
partitions and Poisson coefficients (which \emph{a priori} 
do not have an obvious connection).  

We remark as well that the form of the ratios in both 
algorithms support the theoretical basis of staircase 
scaling \cite{Englert:2011cg,Gerwick:2011tm,Gerwick:2012hq} as 
the ratios formally go to a non-zero constant in the large multiplicity 
limit (and not to $0$).
\smallskip

\underline{Relation to parton shower:}  
In this section we define precisely how our coefficients in the 
exclusive-$k_t$ (Durham) algorithm relate to final state parton 
shower generation.  
We start by considering the evolution along a single quark 
line, implemented in a simple Sudakov veto algorithm at 
fixed coupling \cite{Seymour:1994df,Sjostrand:2006za}.  
Repeating this algorithm a large number of 
times, we find as expected that the splitting probability 
converges to
\begin{equation}
P_1 = 1 - \exp(- \mathcal{P}_1).
\end{equation}
where 
\begin{equation}
\mathcal{P}_1 = \frac{\alpha_s}{2 \pi} C_F \log^2\left(\frac{Q}{Q_0}\right)
\end{equation}
in the limit $Q/Q_0 $ large, and also for $\mathcal{P}_1$ large.  
This process generates a distribution of splitting scales for the first emission 
$Q_1$, which we take as the starting scale for a new Sudakov 
veto process in every other way identical to the first.   The emission 
probability of the second process is  
\begin{equation}
\mathcal{P}_2^{\text{correlated}} = 1 - \exp(- \mathcal{P}_2^{\text{correlated}}).
\end{equation}
In the limit where $Q \gg Q_0$ but $\mathcal{P}_2^{\text{correlated}}\ll \mathcal{P}_1  \ll 1$, 
we can relate $\mathcal{P}_2^{\text{correlated}}$ to the \emph{original} 
external scales $Q$ and $Q_0$ of the process through 
\begin{equation}
P_2^{\text{correlated}}= c^{(2)}_1\frac{\alpha_s}{2 \pi} C_F \log^2\left(\frac{Q}{Q_0}\right).
\end{equation}
Here we have included our previously computed resolved 
coefficient $c_1^{(2)} = 1/12$.  Repeating the process of feeding 
in intermediate splitting scales to generate new Sudakov veto processes, it 
is possible to map out our resolved coefficients on arbitrary 
topologies.  For example taking the intermediate scale $Q_2$ generated 
from the second process into a third Sudakov veto process, we generate 
$c_2^{(3)}$.

Having related (and checked numerically) this stripped-down form 
of the parton shower with our analytic coefficients, it is necessary to 
make a few comments on their interplay:

\begin{itemize}
\item
One can imagine generating an exclusive parton final state 
efficiently by fixing the emission history first.  The emission 
history is chosen from the distribution of primordial coefficients 
generated in our algorithm (which can always be put into 
1-to-1 correspondence with a splitting history).

\item
With the emission 
history in place the problem of fully reconstructing the partonic 
final state is recast as the generation of intermediate 
kinematics.  In principle, this could proceed using standard 
forward evolution, where the $k_t$ and partonic energy 
fractions $zE$ and $(1-z)E$ are chosen locally 
from the distribution of corresponding unintegrated Altarelli-Parisi 
splitting functions \cite{Ellis:1991qj}.  However, the most simple 
implementation would certainly lead to a large unweighing 
efficiency, as emissions further down the parton tree would fail 
basic kinematic vetos.  
Assigning kinematics efficiently to a predetermined splitting 
history may be possible in a backward evolution picture, 
though again, some thought is needed to avoid large inefficiencies.  

\item
We speculate that efficiently relating coefficients of splitting histories  
to the distribution of shower paths arriving at a specific phase space 
point could be useful to programs reliant on the latter.   
This may be applicable for shower deconstruction \cite{Soper:2011cr} 
and perhaps more involved shower algorithms \ie 
Ref.~\cite{LopezVillarejo:2011ap}.  
\end{itemize}

\underline{Exclusive jet rate calculations:}  A further application 
for our coefficients is the promotion of potentially large multiplicity fixed order 
QCD calculations to higher levels of exclusiveness.  This is done 
by matching with our (analytic) unresolved coefficients although 
one should bear in mind that NLO + NLL accuracy cannot be 
achieved without including the wide angle soft gluon component, 
which is process specific and not contained in the coherent branching 
formalism.  Therefore we claim that such an NLO matched calculation 
contains only LO resummed accuracy in the large logarithmic 
regime.  That being said, this opens up the possibility for example 
for an analytic calculation of a 5 jet rate which is physically sensible 
in all regions of phase space.

\underline{Extension to next-to-leading logarithms:}  Although 
we have established the connection between our analytic coefficients 
and a stripped-down parton shower via interleaved Sudakov veto 
algorithms, it is not possible to make strong claims between to the 
full parton shower without accounting for the additional effects 
solely present in the shower.  
The specific pieces include the finite part to the splitting 
functions, the $g\to q\bar{q}$ splitting function, the running 
coupling and the full kinematics.  The first is trivially implemented while 
our ``color stripped" definitions of the coefficients make the second 
approachable.  The inclusion of the running coupling 
requires new steps, but again there is a well defined pattern 
here which is repeated.  It remains to be seen whether these 
additions render the recursive formula prohibitively complicated.
\smallskip

\underline{Relation to singularity structure of gauge theories:}
Our recursive construction offers a connection to the structure 
of all-order singularities in gauge theory amplitudes.  It was 
shown some time ago in Ref.~\cite{Gatheral:1983cz} 
that in the eikonal limit the maximally non-abelian graphs 
exponentiate and therefore enumerate the complete singularity 
structure\footnote{The name 'maximally non-ablelian" was replaced  
by 'color connectedness' in Ref. \cite{Frenkel:1984pz}, which became the 
literature standard.  For our purposes, since in this formalism we do 
not account for non-trivial color correlating virtual gluons, maximally 
non-abelian suits just fine.}.  Schematically this formula in the notation 
of Ref.~\cite{Gatheral:1983cz} 
reads 
\begin{equation}
F_r=\exp\left[ \sum_{s=0}^{\infty} \tilde{F}_s\right]_{r} ,
\label{gath_for}
\end{equation}
where the squared amplitude for $r$ gluons indicates that 
we keep this order in the expansion of the exponential on the 
RHS and $\tilde{F}_s$ is the maximally non-abelian contribution.
Since our resolved coefficients represent some part of  
the leading singularities in the $L\to \infty$ sense, it is no 
surprise that we find a similar construction.  Noting that at 
the level of the (leading) logarithms
\begin{alignat}{5}
\tilde{F}_{n+1} \sim \sum_{j=0}^{n-1} c^{(n)}_{j}  d^{(n)} 
= c^{(n+1)}_n
\label{subs}
\end{alignat}
\eq{final} is precisely the expanded interpretation of 
\eq{gath_for}.  Upon inserting 
\eq{subs} all of the coefficients appearing on the RHS of 
the recursive formula arise from maximally non-abelian splitting 
histories.      
Note the crucial role played by the symmetry factor $S$, 
which embodies the various terms in the expanded exponentials.

Using the substitution \eq{subs}, the dependence 
on the jet algorithm, previously entering only via $d^{(n)}$, 
completely disappears.  Thus our prescription ultimately 
amounts to merely dividing over the identical boson phase 
space.  In particular, all steps (other than the 1st moment) 
are identical between the two classes of jet algorithms 
considered, and we speculate that this holds for any 
exponentiating IR safe regulator.

Eq. \eq{gath_for} becomes exact in the strict eikonal 
limit.  Similarly, we can define the precise limit of 
\eq{final} when the resolved coefficients define the 
exact cross-section.

\begin{itemize}
\item $N_C \to \infty$. 
\item $ \alpha_S\to 0 $ and $L \to \infty$ 
with $\alpha_S L^2$ held fixed 
and $\ll1$.
\end{itemize}

The first point restates the fact that any formalism built on $1\to 2$ 
splittings is inherently leading order in color.  The small coupling and large 
logarithm in the second condition eliminate the contribution from 
the sub-leading logarithms, the finite terms and the unresolved 
double leading logarithms.  An alternative statement for the second 
condition in terms of physical scales is $Q,Q_0\to \infty$ with 
$Q \gg Q_0$.   Although this is an idealized limit for QCD in 
realistic collider processes it is increasingly justified with higher 
energy due to asymptotic freedom.

\section{Conclusions}

In this work we presented a recursive formula for computing 
logarithmic coefficients in QCD.  These were used to study 
the high multiplicity behavior of gluon rates.  However, the main 
interest in these coefficients is that they arise from the same 
physics as a parton shower, and therefore provide an additional 
handle for comparison.  As the LHC provides more high precision 
QCD intensive date, it is becoming more urgent to assess our 
reliance on parton shower Monte Carlo, and we hope that providing  
a better analytic understanding will assist with this task.

\acknowledgments

I thank Steffen Schumann and Bryan Webber for comments 
on the draft, as well as for the collaboration which inspired this project.  
In addition I thank Peter Schichtel for numerous conversations on the 
subject.  Diagrams in this paper were created with JaxoDraw \cite{Binosi:2003yf}.  
I acknowledge support by the Bundesministerium f\"ur Bildung und 
Forschung under contract 05H2012.

\begin{appendix}



%

\section{Resummed rates from generating functional}
One of the checks on our method is direct comparison to the jet 
rates from the straight-forwards computation within the generating 
functional formalism.  We list here the resummed $6$ jet fraction.  
Lower multiplicities can be found elsewhere \cite{Catani:1991hj,Leder:1996py}. 
Defining
{\small
\begin{equation}
\Delta_j(t) 
= \exp \left[ -\int_{Q_0^2}^{t} dt' \; \Gamma_j(t,t') \right] \, .
\label{eq:sudakov_def}
\end{equation}
\begin{alignat}{5}
\Gamma_j (t,t')
&\; = \;C_j \,\frac{\alpha_s(t')}{2\pi t'}
\left( \log \frac{t}{t'} - A_j \right) \; ,
\label{eq:splitting}
\end{alignat}
where $j=q,g$ we find

\allowdisplaybreaks
\begin{alignat}{5}
\label{f6}
f_6 &= [\Delta_q(Q)]^2 \left[ \frac{2}{3} 
\left( \int_{Q_0}^Q dq \,\Gamma_q(Q,q) \Delta_g (q) \right)^4 \right.\notag \\
&+ \;\;
4 \left( \int_{Q_0}^Q dq \,\Gamma_q(Q,q) \Delta_g (q)\right)^2
\left( \int_{Q_0}^Q dq \,\Gamma_q(Q,q) \Delta_g (q) 
\int_{Q_0}^q dq' \,\Gamma_g(q,q') \Delta_g (q')\right) \notag \\
&+ \;\;
4 \left( \int_{Q_0}^Q dq \,\Gamma_q(Q,q) \Delta_g (q)\right) \notag \\ 
& \quad \times \left( \int_{Q_0}^Q dq \,\Gamma_q(Q,q) \Delta_g (q) 
\int_{Q_0}^q dq' \,\Gamma_g(q,q') \Delta_g (q')
\int_{Q_0}^{q'} dq'' \,\Gamma_g(q',q'') \Delta_g (q'')\right) \notag \\
&+ \;\;
2 \left( \int_{Q_0}^Q dq \,\Gamma_q(Q,q) \Delta_g (q)\right)
\int_{Q_0}^Q dq \,\Gamma_q(Q,q) \Delta_g (q) 
\left(\int_{Q_0}^q dq' \,\Gamma_g(q,q') \Delta_g (q')\right)^2 \notag \\
&+ \;\;
2 \left( \int_{Q_0}^Q dq \, \Gamma_{q}(Q,q) \Delta_g (q) 
\int_{Q_0}^q dq' \, \Gamma_g(q,q')  \Delta_g (q')\right)^2
\notag \\
&+ \;\;
\frac{1}{3} \int_{Q_0}^Q dq \,\Gamma_q(Q,q) \Delta_g (q)
\left( \int_{Q_0}^q dq' \,\Gamma_g(q,q') \Delta_g (q') \right)^3 \notag \\
&+ \;\;
2 \int_{Q_0}^Q dq \,\Gamma_q(Q,q) \Delta_g (q)
\left( \int_{Q_0}^q dq' \,\Gamma_g(q,q') \Delta_g (q')\right) 
\notag \\
&\quad \times \left( \int_{Q_0}^q dq' \,\Gamma_g(q,q') \Delta_g (q')
\int_{Q_0}^{q'} dq'' \,\Gamma_g(q',q'') \Delta_g (q'') \right)
\notag \\
&+ \;\;
\int_{Q_0}^Q dq \,\Gamma_q(Q,q) \Delta_g (q)
\left( \int_{Q_0}^q dq' \,\Gamma_g(q,q') \Delta_g (q')\right) 
\left( \int_{Q_0}^{q'} dq'' \,\Gamma_g(q',q'') \Delta_g (q'') \right)^2 
\notag \\
&+ \;\;
2 \int_{Q_0}^Q dq \,\Gamma_q(Q,q) \Delta_g (q)
\notag \\ 
& \quad \times \left. \left( \int_{Q_0}^q dq' \,\Gamma_g(q,q') \Delta_g (q')
\int_{Q_0}^{q'} dq'' \,\Gamma_g(q',q'') \Delta_g (q'') 
\int_{Q_0}^{q''} dq''' \,\Gamma_g(q'',q''') \Delta_g (q''')
\right) \right].
\end{alignat}}

\end{appendix}


\end{document}